\documentclass[twocolumn,secnumarabic,amssymb, nobibnotes, aps, prd]{revtex4-1}

\usepackage[caption=false]{subfig}
\usepackage{amssymb}
\usepackage{amsmath}
\usepackage{commath}
\usepackage{graphicx,bm}
\usepackage{verbatim}
\usepackage{graphicx}
\usepackage{dcolumn}
\usepackage{bm}
\usepackage{multirow}
\usepackage{booktabs}
\usepackage{chngpage}
\usepackage{float}
\usepackage{setspace}

\newcommand{\bef}{\begin{figure}}
\newcommand{\eef}{\end{figure}}
\newcommand{\bec}{\begin{center}}
\newcommand{\eec}{\end{center}}
\newcommand{\be}{\begin{equation}}
\newcommand{\ee}{\end{equation}}

\begin{document}

\title{Evolving extrinsic curvature and the cosmological constant problem}%

\author{A.J.S. Capistrano}%
\email[]{abraao.capistrano@unila.edu.br}
\affiliation{1 Federal University of Latin-American Integration, 85867-970, P.o.b: 2123, Foz do Igua\c{c}u-PR, Brazil}
\affiliation{2 Casimiro Montenegro Filho Astronomy Center, Itaipu Technological Park, 85867-900, Foz do Iguassu-PR, Brazil}

\author{L.A. Cabral}%
\email[]{cabral@uft.edu.br}
\affiliation{3 Federal University of Tocantins, 77804-970, Aragua\'{i}na-TO, Brazil}

\date{\today}

\begin{abstract}
The concept of   smooth   deformation  of  Riemannian manifolds  associated  with  the  extrinsic  curvature is explained and applied to the FLRW cosmology. We  show  that such deformation  can be   derived  from Einstein-Hilbert-like  dynamical   principle  producing an observable  effect in the  sense of  Noether. As a result, we notice on how the extrinsic curvature compensates both quantitative and qualitative difference between the cosmological constant $ \Lambda$ and the vacuum energy $\rho_{vac}$ obtaining the observed upper bound for the cosmological constant problem at electroweak scale. The topological characteristics of the extrinsic curvature are discussed showing that the produced extrinsic scalar curvature is an evolving dynamical quantity.
\end{abstract}

\maketitle

\section{Introduction}
In  a previous  investigation \citep{GDEI} we studied a modification imposed on  Friedman's  equation   when the  standard model of the  universe
is    regarded  as  an  embedded   space-time \citep{QBW}. It was shown that a more fundamental  explanation  for the  dynamics of the extrinsic
curvature  is  required and given by the Gupta equations \citep{Gupta}.  As a result, the accelerated expansion of the universe  could  be  explained as an effect of the  extrinsic  curvature.

In this work, we study the cosmological constant (CC) problem that primarily consists in a seemingly unexplainable difference between the small value of the CC estimated by cosmological observations to be $\Lambda/8\pi G \sim 10^{-47}\;GeV^4$ and the theoretical value is given by the vacuum energy density that results from gravitationally coupled quantum fields in space-time estimated to be of the order of $< \rho_{v}> \sim 10^{71}\;GeV^4$. Such large difference cannot be eliminated by renormalization techniques in quantum field theory as it would require an extreme fine tuning \citep{Zeldowich,weinberg}. In the last decade, it became a central issue in the context of the $\Lambda$CDM cosmological model regarded as the simplest model for the accelerated expansion of the universe. In addition, another dilemma requires attention that is a proper explanation to the apparently coincidence between the current matter density energy and CC (as interpreted as the vacuum energy) commonly known as the \emph{coincidence problem} \citep{carroll,weinberg2}. A varieties of solutions for the CC problems have been proposed in literature such as in general relativity \citep{3,4,5}, strings \citep{6} and branes \citep{7,8,9}, conformal symmetry of gravity \citep{cadoni} and other works \citep{steinhardt, shapiro1,shapiro2,simone,raul}.

In a different direction, we address the CC problem from a geometrical approach. We use essentially that in embedded space-time the gauge fields remain confined to the embedded space but the gravitational field propagates along the extra dimensions (similar to the brane-world program originally proposed in \cite{ADD}). On the other hand, it is important to point out that the difference between the vacuum energy and the CC is hidden in most brane-world models because the extrinsic curvature is commonly replaced by a function of the confined source fields. As commonly thought, the only accepted relation of the extrinsic  curvature   with matter sources is  the Israel-Lanczos  boundary    condition,  as  applied  to the Randall-Sundrum brane-world cosmology \citep{RS,RS1}.  However,  this condition fixes  once for  all   the extrinsic  curvature  and  does not  follow  the  dynamics of the brane-world. Other approaches  have been developed with no need of particular junction conditions \citep{sepangi,sepangi1} and/or with different junction conditions which lead to several approaches of brane-world models widely studied in literature \citep{maeda,maartens,anderson,sahni,Tsujikawa,gong}.

The main purpose of this paper is to show that the CC problem comes from a fundamental origin, not only because it involves the structure of the Einstein-Hilbert principle but also because it reinforces a clearly distinction between of gravitation from gauge fields. In what follows, we focus on the CC problem at low redshift, since it is verified in the present epoch \citep{shapiro1,shapiro2}. We obtain an explicit relation involving the extrinsic curvature and the absolute difference between  CC and the vacuum energy density. As we shall see, we show that the dynamics of the extrinsic curvature has a more profound meaning which a four-dimensional observer can detect a difference between Einstein's CC and the confined vacuum energy through a conserved quantity. Another relevant aspect is  how the extrinsic scalar $Q$ evolves and its topological consequences. In this framework we are neglecting fluctuations and/or effects of structure formation. Finally, remarks are presented in the conclusion section.

\section{The FLRW embedded universe}

\subsection{Modified friedmann equations}
We start with the Friedmann-Lema\^{\i}tre-Robertson-Walker (FLRW) line  element
in coordinates $(r,\theta,\phi,t)$  given by
\begin{equation}\label{eq:line element}
ds^2=\;-dt^2+a^2\left[dr^2+f_{\kappa}^2(r)\left(d\theta^2+\sin^2\theta
d\varphi^2\right)\right]\;,
\end{equation}
where $f_{\kappa}(r)=\sin r$, $r$,$\sinh r$ correspond to   $\kappa$ = (1,
0, -1), and the term $a=a(t)$ is the expansion parameter. This  model  can be regarded as a four-dimensional hypersurface
dynamically evolving  in  a  five-dimensional ``bulk'' with constant
curvature. This geometry induced by four-dimensional FLRW line element is completely embedded in a five-dimensional bulk. The Riemann tensor is given by \footnote{We  use the same notation and  conventions as in \cite{GDEI}. For the present application in five dimensions, capital Latin indices run from 1 to 5. Small case Latin indices refer to the extra dimensions only. In the present case, we have only one extra dimension. All Greek indices refer to the embedded space-time counting from 1 to 4.}
$$\mathcal{R}_{ABCD}=
K_*\left(\mathcal{G}_{AC}\mathcal{G}_{DB}-\mathcal{G}_{AD}\mathcal{G}_{CB}\right),
$$
where $\mathcal{G}_{AB}$ denotes the bulk metric components in
arbitrary coordinates. The  constant  curvature $K_{\ast}$ has three possible values : it is
either zero (flat bulk), a positive (de Sitter) or
negative (anti-de Sitter) constant curvatures.

Since we are dealing with embedding of geometries, the  general  solution  was  given by  John  Nash in 1956 \cite{Nash},  using only differentiable (non-analytic) properties. In  short, starting with an  embedded Riemannian  manifold  with    metric $g_{\mu\nu}$  and  extrinsic  curvature $k_{\mu\nu}$, Nash  showed  that any other  embedded  Riemannian geometry can  be  generated  by differentiable perturbations,  with metric $\tilde{g}_{\mu\nu}= g_{\mu\nu} + \delta  g_{\mu\nu} $,  where
\be
\delta g_{\mu\nu} =-2k_{\mu\nu a}\delta y^a\;, \label{eq:York}
\ee
and  where  $\delta y^a$ is an infinitesimal displacement in one of the  extra dimension. From this  new metric, we obtain a new extrinsic curvature $k_{\mu\nu }$  and the  procedure  can  be repeated  indefinitely
\be
g_{\mu\nu}  =  g_{\mu\nu}  +  \delta y^a \, k_{\mu\nu  a}  +
\delta y^a\delta y^b\, g^{\rho\sigma}
k_{\mu\rho a}k_{\nu\sigma b}\cdots  \label{eq:pertu}
\ee
For this reason, depending on the size of the bulk, the embedding map can be well defined and, at first, one does not need to perturb the line element in eq.(\ref{eq:line element}) in the \emph{y}-coordinate  direction, since Nash theorem already guarantees this property.

Taking the perfect fluid of the  standard cosmology  as composed of  ordinary matter
interacting  with  gauge  fields, it must  remain   confined
to  the   four-dimensional space-time  on all stages of the
evolution  of the universe. Since all  cosmological observations
point  to  an accelerated expanding   universe  towards  a  de Sitter
configuration \citep{Perlmutter,Riess},   we  choose   $K_*>0$,
although  our  results  also hold for  any  other   choice of
$K_*$. The  bulk geometry is  actually  defined by the
Einstein-Hilbert principle that  leads  to the Einstein
equations
\be
{\mathcal R}_{AB} -\frac{1}{2} {\mathcal R}  {\mathcal G}_{AB}=\alpha_*  T^*_{AB}\;.
\label{eq:BE0}
\ee

The   confinement  condition   implies that     $K_* = \Lambda /6$ and
$T^*_{AB}  $  denotes the energy-momentum  tensor  of the known  sources.

The  confinement  of  gauge  fields  and  ordinary  matter  are  a  standard  assumption specially in what concerns the brane-world program  as a part  of  the  solution of the hierarchy problem  of the fundamental  interactions:  the  four-dimensionality of     space-time  is  a  consequence of the invariance of Maxwell's  equations  under  the Poincar\'{e}  group. Such   condition  was  latter extended  to  all  gauge  fields expressed  in terms  of  differential forms  and  their  duals. However, in  spite of  many  attempts,  gravitation, in the sense of  Einstein,  does not  fit in  such  scheme.   Thus,   while  all  known  gauge  fields  are  confined to the four-dimensional   submanifold,  gravitation  as defined   in the  whole  bulk  space by  the Einstein-Hilbert principle, propagates in the  bulk.  The   proposed solution of the  hierarchy problem    says  that  gravitational  energy  scale  is    somewhere  within TeV  scale.

The  most general  expression  of this  confinement is  that  the  confined    components of  $T_{AB}$  are
proportional  to   the     energy-momentum tensor  of general relativity:  $\alpha_* T_{\mu\nu}= -8\pi G T_{\mu\nu}$. On the other  hand,   since only gravity propagates  in the  bulk   we  have  $T_{\mu a}=0$ and  $T_{ab}=0$.

Since we are dealing with the relations between embedded space-times, we can restrict our analysis to the local embedding of five-dimensions which can be summarized defining an embedding map $\mathcal{Z}:\bar{V}_{4}\rightarrow V_{5}$ admitting that $\mathcal{Z}^{\mu}$ is a regular and differentiable map with $V_{4}$ and $V_5$ being the embedded space-time and the bulk, respectively. The components $\mathcal{Z}^{A} =\;f^{A}(x^{1},...,x^{4})$ associate with each point of $V_{4}$ to a point in $V_{5}$ with coordinates $\mathcal{Z}^{A}$ that are the components of the tangent vectors of $V_{4}$. Moreover, taking the tangent, vector and scalar components of eq.(\ref{eq:BE0}) defined in the Gaussian frame vielbein $\{\mathcal{Z}^A_{,\mu},\eta^A\}$, where $\eta^{A}$ are the components of the normal vectors of $V_{4}$, one can write the set of equations in the  five-dimensional de Sitter bulk \citep{GDEI,QBW}
\begin{eqnarray}\label{eq:BE1}
&&R_{\mu\nu}-\frac{1}{2}Rg_{\mu\nu}+\Lambda g_{\mu\nu}-Q_{\mu\nu}=
-8\pi G T_{\mu\nu}\;, \hspace{4mm}\\
&&k_{\mu;\;\rho}^{\;\rho}-h_{,\mu} =0\;,\label{eq:BE2}
\end{eqnarray}
where  $T_{\mu\nu}$ is the 4-dimensional energy-momentum tensor of
the perfect  fluid  expressed  in  co-moving coordinates as
$$T_{\mu\nu}=(p+\rho)U_{\mu}U_{\nu}+p\;g_{\mu\nu},\;\;\;U_{\mu}=\delta_{\mu}^{4}\;.$$

It is important to point out that the quantity $Q_{\mu\nu}$  is a completely geometrical term  given
by
\begin{equation}\label{eq:qmunu}
  Q_{\mu\nu}=g^{\rho\sigma}k_{\mu\rho }k_{\nu\sigma}- k_{\mu\nu }h -\frac{1}{2}\left(K^2-h^2\right)g_{\mu\nu}\;,
\end{equation}
where $h= g^{\mu\nu}k_{\mu\nu}$, $h^2=h.h$ and $K^{2}=k^{\mu\nu}k_{\mu\nu}$. It
follows  that  $Q_{\mu\nu}$  is  conserved in the sense  that
\begin{equation}\label{eq:cons}
  Q^{\mu\nu}{}_{;\nu}=0\;.
\end{equation}

The  general  solution  for  eq.(\ref{eq:BE2}) using the FLRW  metric is
\begin{eqnarray}
 &&k_{ij}=\frac{b}{a^2}g_{ij},\;\;
k_{44}=\frac{-1}{\dot{a}}\frac{d}{dt}\frac{b}{a}\nonumber
\end{eqnarray}
in this case $i, j= 1, 2, 3$, where  we also notice that  the ``warping'' function    $b(t)=k_{11}$  remains   an
arbitrary function of time.  This  follows  from the confinement of the   gauge  fields that  produces  the  homogeneous  equation in eq.(\ref{eq:BE2}).

The usual  Hubble parameter in terms of the expansion scaling factor $a(t)=a$ is denoted by $ H = \dot{a}/a$  and the
extrinsic parameter    $B= \dot{b}/b$. Solving the set of eq.(\ref{eq:BE1}) and eq.(\ref{eq:BE2}), one can obtain
\begin{eqnarray}
 &&
 k_{44}=-\frac{b}{a^{2}}\left(\frac{B}{H}-1\right)g_{44},\; h=\frac{b}{a^2}\left(\frac {B}{H}+2\right)\label{eq:hk},  \\
&&K^{2}=\frac{b^2}{a^4}\left( \frac{B^2}{H^2}-2\frac{B}{H}+4\right),\\
&&Q_{ij}= \frac{b^{2}}{a^{4}}\left( 2\frac{B}{H}-1\right)
g_{ij},\;Q_{44} = -\frac{3b^{2}}{a^{4}},
  \label{eq:Qab}\\
&&Q= -(K^2 -h^2) =\frac{6b^{2}}{a^{4}} \frac{B}{H}\;. \label{Q}
 \end{eqnarray}
In the case of eq.(\ref{eq:Qab}), consider $i, j= 1, 2, 3$.

Replacing these results in  eq.(\ref{eq:BE1}), we obtain the Friedman equation modified  by the  extrinsic  curvature as
\begin{equation}\label{eq:Friedman}
\left(\frac{\dot{a}}{a}\right)^2+\frac{\kappa}{a^2}=\frac{4}{3}\pi
G\rho+\frac{\Lambda}{3}+\frac{b^2}{a^4}\;\;.
\end{equation}

\subsection{Gupta extrinsic equation and the unique solution for the ``warping'' function b(t)}
The  arbitrariness   of  ``warping'' function $b(t)$ is  a  consequence  of the homogeneity of  eq.(\ref{eq:BE2}) which follows  from the  confinement  condition   $T^*_{\mu a}=0$.  If   these  components  were  non  zero, we  would   violate the   intended  solution of the hierarchy  problem. For instance, the  Randall-Sundrum  brane-world  models avoid  such  difficulty  by   fixing the brane-world as  a  boundary  at   $y=0$ and applying the   Israel-Lanczos boundary  condition. In order to obtain dynamical equations, the Gupta equations were used and the function  $b(t)$  was  determined   by constructing the dynamics of  extrinsic curvature $k_{\mu\nu}$ interpreted as a component of gravitational field besides the metric $g_{\mu\nu}$.

In short, the study of linear massless spin-2 fields in Minkowski space-time  by  Fierz  and  Pauli dates back to late 1930's~\cite{FierzPauli}. In 1954, Gupta \cite{Gupta} noted that the Fierz-Pauli equation has a remarkable resemblance with the linear approximation of Einstein's equations for the gravitational field, suggesting that such equation  could be just the linear approximation of a more general, non-linear equation for massless spin-2 fields. In reality, he  found that any spin-2 field in Minkowski space-time must satisfy an equation that has the same formal structure as Einstein's equations. This amounts to saying that, in the same way as Einstein's  equations can be obtained by an infinite sequence of infinitesimal perturbations of the linear gravitational equation, it is possible to obtain  a non-linear equation for any spin-2 field by applying an infinite sequence of infinitesimal perturbations to the Fierz-Pauli equations. The result obtained by  S. Gupta is an Einstein-like system of equations \cite{Fronsdal,Gupta}.

In the  following  we   use an analogy with the derivation of the Riemann  tensor  to  write  Gupta's  equation in a Riemannian  manifold  with metric geometry   $g_{\mu\nu}$  embedded in  a five-dimensional bulk. In this analogy we construct a  ``connection'' associated with $k_{\mu\nu}$  and then,  find its  corresponding Riemann tensor, but keeping in mind that the geometry of the embedded space-time has been  previously   defined by $g_{\mu\nu}$. To do so, we define  the  tensors
\be
f_{\mu\nu} = \frac{2}{K}k_{\mu\nu}, \;\; \mbox{and}
\;\;f^{\mu\nu} = \frac{2}{K}k^{\mu\nu}\;,
\label{eq:fmunu}
\ee
 so that $f^{\mu\rho}f_{\rho\nu} =\delta^\mu_\nu$. In the sequence we construct the ``Levi-Civita  connection'' associated with $f_{\mu\nu}$, based on the analogy with  the ``metricity condition''  $f_{\mu\nu||\rho}=0$,  where $||$ denotes  the covariant derivative with respect to $f_{\mu\nu}$ (while keeping the usual $(;)$ notation for the covariant derivative with respect to $g_{\mu\nu}$).  With this condition  we obtain the  ``f-connection''
$$
\Upsilon_{\mu\nu\sigma}=\;\frac{1}{2}\left(\partial_\mu\; f_{\sigma\nu}+ \partial_\nu\;f_{\sigma\mu} -\partial_\sigma\;f_{\mu\nu}\right)
$$
and
$$
\Upsilon_{\mu\nu}{}^{\lambda}= f^{\lambda\sigma}\;\Upsilon_{\mu\nu\sigma}
$$
and  the ``f-Riemann tensor''
$$
\mathcal{F}_{\nu\mu\alpha\lambda}= \;\partial_{\alpha}\Upsilon_{\mu\lambda\nu}- \;\partial_{\lambda}\Upsilon_{\mu\alpha\nu}+ \Upsilon_{\alpha\sigma\mu}\Upsilon_{\lambda\nu}^{\sigma} -\Upsilon_{\lambda\sigma\mu}\Upsilon_{\alpha\nu}^{\sigma}
$$
and the ``f-Ricci tensor'' and the ``f-Ricci scalar'', defined with $f_{\mu\nu}$ are, respectively,
$$
{\cal F}_{\mu\nu} =  f^{\alpha\lambda}\mathcal{F}_{\nu
\alpha\lambda\mu}
\;\;\mbox{and}\;\;\mathcal{F}=f^{\mu\nu}\mathcal{F}_{\mu\nu}
$$
Finally, write  the  Gupta equations for the $f_{\mu\nu}$ field
\begin{equation}
\label{eq:gupta}
\mathcal{F}_{\mu\nu}-\frac{1}{2}\mathcal{F} f_{\mu\nu} =\;\alpha_f\tau_{\mu\nu}
\end{equation}
where $\tau_{\mu\nu}$ stands for the source of the f-field, with coupling constant $\alpha_f$. However, unlike the case of  Einstein's  equations,  here we do not  have the  equivalent  to  the Newtonian weak field limit, then we  cannot tell  about   the  nature of the source  term  $\tau_{\mu\nu}$  based on  experience. As  a  first guess  we may start with the  simplest  ``f-Ricci-flat''  equation
\begin{equation}
\label{eq:guptaflat} \mathcal{F}_{\mu\nu}=0\;.
\end{equation}

As we use eq.(\ref{eq:line element}) and calculate eq.(\ref{eq:guptaflat}), we can take eq.(\ref{eq:qmunu}) and eq.(\ref{eq:cons}) and obtain the expression for the ``warping'' function $b(t)$ that is given by
\begin{equation}\label{eq:geb}
b(t)= \alpha_0a^{\beta_0}e^{\pm
\frac{1}{2}\gamma(t)}\;.
\end{equation}
We denote $\alpha_0 = b_0/ a_0^{\beta_0}$, $a_0$ by the present value of the expansion scaling factor and $b_0$ is an integration constant representing the current warp of the universe. Also, the exponential function has the exponent given by $\gamma(t)=\sqrt{4\eta_0a^4 - 3}-\sqrt{3}\arctan\left(\frac{\sqrt{3}}{3}\sqrt{4\eta_0a^4 -3}\right)$. The two signs represent two possible signatures of the evolution of the function $b(t)$ which can be important to study how it evolves, and in the next section, we study how it can be related to the CC problem.

\section{The balance through extrinsic curvature}
Analyzing eq.(\ref{eq:BE1}),  a four-d observer realizes that the quantum vacuum energy density $<\rho_v>$ can be related to $\Lambda g_{\mu\nu}- Q_{\mu\nu}$ different from the case of general relativity that we only have the term $\Lambda g_{\mu\nu}$. Thus, taking eq.(\ref{eq:BE1}), we have
$$R_{\mu\nu}-\frac{1}{2}Rg_{\mu\nu}+\Lambda g_{\mu\nu}-Q_{\mu\nu}=-8\pi GT_{\mu\nu}\;\;,$$
and considering the vacuum contribution $T_{\mu\nu}= - <\rho_v> g_{\mu\nu}$, one can write
\be
8\pi G<\rho_v> g_{\mu\nu} = \Lambda g_{\mu\nu} - Q_{\mu\nu}\;\;,
\ee
and contracting with $g_{\mu\nu}$, we obtain
\be
<\rho_v>- \rho_{\Lambda} = -\frac{Q}{32\pi G}
\ee
where $Q=\; g^{\mu\nu}Q_{\mu\nu}$ is the trace of $Q_{\mu\nu}$. In addition, using eq.(\ref{Q}) we can write
\be\label{eq:relação Q e lambda e rho1}
<\rho_v>- \rho_{\Lambda} = -\frac{6b^{2}B}{32\pi G a^{4}H}\;,
\ee
which indicates that the discrepancy ceases to be if the extrinsic curvature can compensate such difference.

We can make now an analysis of eq.(\ref{eq:relação Q e lambda e rho1}) starting from the ``warping'' function $b(t)$. Thus, we seek a general relation that can relate the expansion parameter $a(t)$ with the difference between $<\rho>_v$ and $\rho_{\Lambda}$. Thus, taking the Gupta solutions in eq.(\ref{eq:geb}) for FLRW cosmology \citep{GDEI} and eq.(\ref{eq:Friedman}), we can write the modified
Friedman equation in terms of the redshift \emph{z} and cosmological parameters as
\begin{equation}\label{eq:modfz}
H^2=H_0^2\left[\Omega_{m}+\Omega_{ext}\;e^{\pm\gamma(z)}\right]\;,
\end{equation}
where $\Omega_{m}$ is the matter density cosmological parameter defined as $\Omega_{m}=\Omega^{0}_{m}(1+z)^3$ and $H_0$ is the current Hubble constant. Hereafter, the upper script ``0'' indicates the present value of certain quantity.

In order to be consistent with \cite{GDEI}, we consider the current value for the expansion factor as $a_0=1$. The term $\gamma(z)$ is written in terms of the redshift \emph{z} and is given by $\gamma(z)=\sqrt{\frac{4\eta_0}{(1+z)^4} - 3}-\sqrt{3}\arctan\left(\frac{\sqrt{3}}{3}\sqrt{\frac{4\eta_0}{(1+z)^4}- 3}\right)$. Inspired by the cosmic fluid analogy as well defined in standard cosmology, the extrinsic cosmological parameter can be written in terms of redshift as $\Omega_{ext}= \Omega^0_{ext} (1+z)^{4-2\beta_0}$ with  $\Omega^0_{ext}= \frac{\alpha^2_0}{H^2_0}$.

Alternatively, we can define  $\Omega_{ext}$ in terms of the extrinsic energy density as
$$\Omega^0_{ext}= \frac{8\pi G }{3}\rho^0_{\;ext}\;,$$
and also the extrinsic energy density as
\begin{equation}\label{eq:extrinsic energy density}
\rho^0_{ext}= \frac{3}{8\pi G} \frac{\alpha_0^2}{H_0^2}\;.
\end{equation}

The parameter $\alpha_0$ can be easily constrained using the normalization $H\rfloor_{z=0} = H_0$. For the present epoch, $\Omega_{total}\rfloor_{z=0} = \Omega_{ext}^0\;\exp(\gamma(0)) + \Omega_m^0 = 1$. Thus, using eq.(\ref{eq:modfz}), one can obtain
\begin{equation}\label{eq:alpha0}
\alpha_0^2= \frac{1- \Omega_m^0}{\exp(\gamma(0))}H_0^2.
\end{equation}
In addition, the estimated value for $\Omega_{ext}^0$ can constrained with the observational values of $\Omega_m^0$ and $H_0$.

Interestingly, using eqs.(\ref{eq:geb}), (\ref{eq:relação Q e lambda e rho1}) and (\ref{eq:extrinsic energy density}), we find
\begin{equation}\label{eq:general expression1}
|<\rho_v>- \rho_{\Lambda}| = \frac{H_0^2}{2} \rho_{ext}\; \xi(z) (1+z)^{4-2\beta_0},
\end{equation}
where
\begin{equation}\label{eq:xi}
\xi(z)=\left(\beta_0 \pm \sqrt{\frac{4\eta_0}{(1+z)^4} - 3}\right) \exp{[\pm\gamma(z)]}.
\end{equation}
From this equation one can obtain that the evolution of the difference of vacuum energy and CC is balanced through the extrinsic curvature evolving on redshift. Hence, in order to test the effectivity of this expression, we calculate that difference for today, i.e, $(z=0)$. To this matter, we use the pair of parameters $(\beta_0, \eta_0)$ that was already constrained in \citep{Capistrano} and adopt the values $\beta_0=2$ and $\eta_0=0.25$. It is important to stress that those values constituted one of the set of solutions (models) that matched the cosmokinetics tests studied in the accelerated expansion of the universe. It was found that $\beta_0$ affects the value of current deceleration parameter $q_0$ and $\eta_0$ rules mainly on the width of the transition phase $z_t$. Based on the fact that eq.(\ref{eq:geb}) can provide different solutions with the term $\gamma$, when it holds for $\pm \gamma(z) = 0$, one can obtain the similar pattern as obtained from phenomenological solutions as shown in \cite{GDEI} that mimics the X-CDM model with a correspondence
\begin{equation}\label{eq:exotic}
4-2\beta_0 = 3(1+ w)\;,
\end{equation}
where the parameter $w$ holds for the exotic X-fluid parameter \citep{turner}. Rather than only reproducing known phenomenological models we are interested also in solutions $\pm \gamma \neq 0$. Thus, we adopt the current value of Hubble constant $H_0$ as $H_0=67.8 \pm 0.9\;km.s^{-1}.Mpc^{-1}$ and  the current matter density parameter $\Omega_m^0 = 0.308 \pm 0.012$ based on the latest observations \citep{Ade}. Thus, using eq.(\ref{eq:extrinsic energy density}), eq.(\ref{eq:general expression1}) turns
\begin{equation}\label{eq:general expression2}
|<\rho_v>- \rho_{\Lambda}| = \frac{3 \alpha_0^2}{16\pi G} \rho^0_{\;ext} \xi(z=0).
\end{equation}
With $H_0 \sim 10^{-42}\;GeV^4$, we apply those values to eq.(\ref{eq:general expression2}) and find that the difference $|<\rho_v>- \rho_{\Lambda}|\lesssim 10^{-47}GeV^4$ matches the upper bound value for the cosmological constant problem in electroweak scale. This shows that the effect of extrinsic curvature has become subtle in the present time. This interpretation seems to be very reasonable since the main process of formation of structures at cosmological scale in universe happened a long time ago and today it appears to be reduced at local scales. Since the extrinsic curvature can warp, bend or stretch a geometry, it is expected that in early times the presence of this perturbational effect played a fundamental role.

\begin{figure*}
\subfloat[]{%
  \includegraphics[height=6cm,width=.49\linewidth]{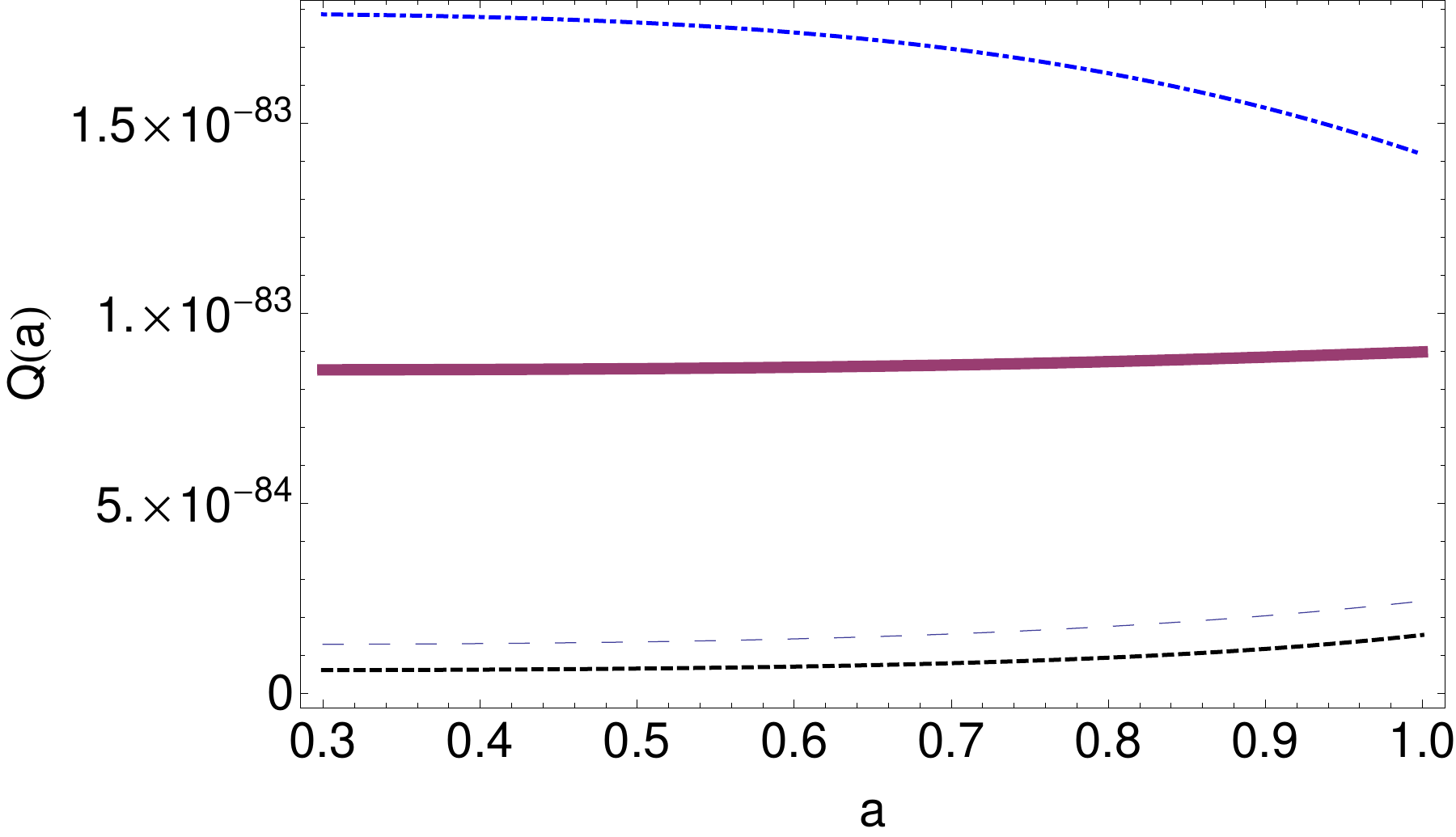}%
}\hfill
\subfloat[]{%
  \includegraphics[height=6cm,width=.49\linewidth]{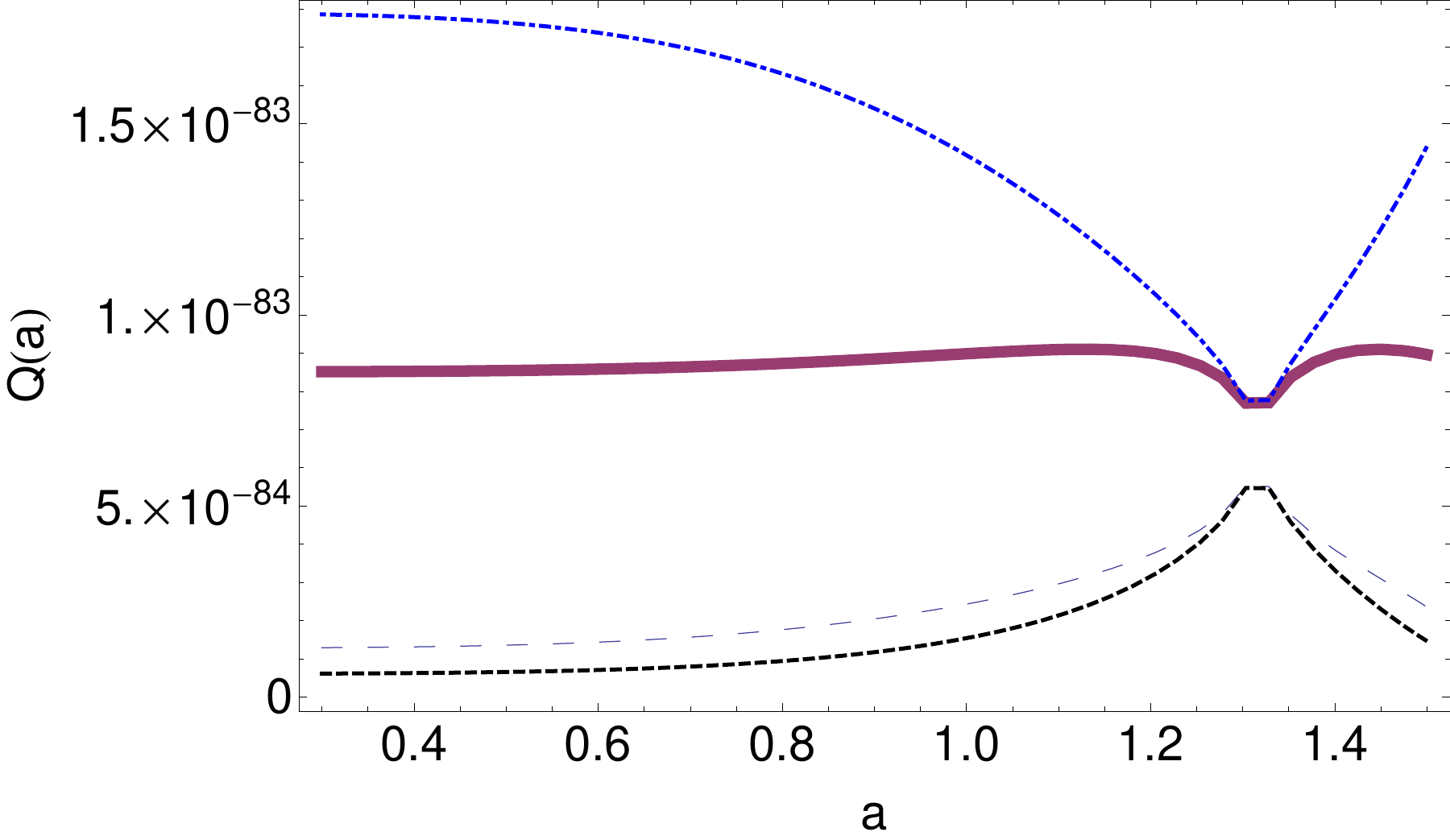}%
}
\caption{In the left figure (a), it is shown the evolution of the absolute value of the extrinsic scalar $Q$ ranging from $a=0.3$ to $a=1$, or equivalently, in redshift $z=0$ to $z=2.3$. The resulting curves are the solutions of different signs of $Q$. In the right figure (b), it is shown an extension of our results until $a=1.5$. Both plots are in logarithm scale.}
\label{fig:qtw}
\end{figure*}

\subsection{The evolving extrinsic scalar}
As already pointed out, the quantity $Q$ is an independent quantity and is defined without the need of existence of $\Lambda$. In order to get an explicit form for the evolution of this quantity, we can estimate how the extrinsic scalar $Q$ evolves as the universe expands. Thus, using the eq.(\ref{eq:geb}), we can rewrite eq.(\ref{Q}) as a function of the expansion factor $a$ in terms of the Hubble constant and the current extrinsic parameter $\Omega_{ext}^0$ as
\begin{equation}\label{eq:Q2}
Q(a) = 6 H_0^2 \Omega_{ext}^0 \xi(a) a^{2\beta_0 -4}.
\end{equation}
One can obtain different solutions that basically depend on the signs from eq.(\ref{eq:xi}) which we use to denote the absolute value of $Q(a)$ as
\begin{eqnarray*}
  Q^{-+} &=&  \varpi(a)\left(\beta_0 - \sqrt{4\eta_0 a^4 - 3}\right)  \exp{[+\gamma(a)]} \; \\
  Q^{+-} &=&  \varpi(a)\left(\beta_0 + \sqrt{4\eta_0 a^4 - 3}\right) \exp{[-\gamma(a)]}\;\\
  Q^{--} &=&  \varpi(a)\left(\beta_0 - \sqrt{4\eta_0 a^4 - 3}\right) \exp{[-\gamma(a)]}\;\\
  Q^{++} &=&  \varpi(a)\left(\beta_0 + \sqrt{4\eta_0 a^4  - 3}\right) \exp{[+\gamma(a)]}\;
\end{eqnarray*}
where we denote $\varpi(a)=6 H_0^2 \Omega_{ext}^0 a^{2\beta_0 -4}$. Moreover, from eq.(\ref{eq:Q2}) one can obtain the resulting plots as shown in the left and right panels in fig.(\ref{fig:qtw}). In both panels, the dashed line represents the solution $Q^{-+}$, the thick line represents the solution $Q^{+-}$, the thick-dashed line represents the solution $Q^{--}$ and the thick-dotted-dashed line represents the solution $Q^{-+}$ and those solutions vary around $10^{-83} \sim 10^{-84} GeV^2$. As shown in the left panel in fig.(\ref{fig:qtw}), we obtain the evolution of the absolute value of the extrinsic scalar $Q$ ranging from $a=0.3$ to $a=1$. In the right panel, we extrapolate the results and they strongly suggest that the solutions presented induce to some changing in topology of an asymptotical future universe. Interestingly, solutions $Q^{++}$ and $Q^{-+}$ present an increasing of the absolute value of $Q$. The former solution provides an earlier acceleration than the latter inducing to a more accelerated regime of the expanding universe, since we are considering the extrinsic curvature the main cause of the accelerated expansion.  On the other hand, $Q^{+-}$ and $Q^{--}$ suggest that after the phase transition at $a\sim 1.3$ the absolute value of $Q$ will decay and both solutions seem to converge to value less than $10^{-84}GeV^2$.

These results reinforce the idea that the extrinsic scalar $Q$ is a dynamical quantity that evolves in time, which is expected for an expanding universe and is roughly of order of the physical CC and the Ricci scalar curvature. Such results have twofold considerations. First, the quantitative issue: the current value of $Q$ is quantitatively similar to the physical cosmological constant and, rather, is a dynamical quantity that dominates the cosmological constant term. Second, the qualitative issues must be account carefully. The extrinsic scalar $Q$ and CC have stinkingly different meanings. The presence of $Q$ show us that CC must be independently from the definition of the source $T_{\mu\nu}$ \citep{maiapinc} and the CC problem has also a topological origin. In this sense, the extrinsic curvature transfers topological characteristics from a Schwarzschild-de Sitter space-time ($\Lambda \neq 0, \kappa \neq 1$) to a Minkowski space-time ($\Lambda \approx 0, \kappa =0$) once Riemannian manifolds also are topological spaces \citep{veb1,veb2}.

Another qualitative aspect refers to the topological and geometrical difference between Minkowski and de Sitter space-times. Those space-times obey different symmetry groups and they are not correlated in the sense that one cannot build a de Sitter space-time starting from a continuous deformation without ripping off the manifold. The lack of a standard reference space-time is a symptom of the Riemann tensor equivalence dilemma and recognized by Riemann himself \citep{Riemann}.

In addition, the In\"{o}nu-Wigner contraction \citep{inonu} tells that we can recover Poincar\'{e} group (ISO(3,1)) from de Sitter group (SO(4,1)) with the limit $\Lambda\rightarrow 0$. However, this is valid for Lie groups due to the fact that they are analytical manifolds. Unfortunately, it does not apply to space-times since they are differentiable manifolds. In this sense, an interesting fact was pointed out in \cite{maiaBMS} that even considering an asymptotically flat space–time ($\Lambda \rightarrow 0$) one does not obtain a Minkowskian flat space but another space-time structure governed by the Bondi-Metzner-Sachs (BMS) group that is a semi-direct product of the Lorentz group with the group of supertranslations. This observation seems to suggest serious constraints on cosmological models and should be investigated in forthcoming studies.

\section{Final remarks}
In a previous paper \citep{GDEI}, we have used a model independent formulation based on the Nash embedding theorem where the
extrinsic curvature is an independent variable required for the definition of the embedding. However, this comes at the price that the extrinsic curvature cannot be completely determined, because Codazzi's equations become homogeneous (incidentally, the Randall-Sundrum model avoids this problem
by imposing the Israel-Lanczos condition on a fixed boundary-like brane-world). Therefore, in order to restore the definition of the extrinsic curvature an additional equation compatible with a dynamically evolving embedded space-time is required. As a rank-2 symmetric tensor, the extrinsic curvature can be seen as a spin-2 field which satisfies Einstein-like equations constituting the so-called Gupta equations for the extrinsic curvature.

The present paper complements that result where the solution of these equations describes not only on how the universe presents an accelerated expansion but also on how it is inner related to the CC problem. At the present cosmic scale, we have shown that the extrinsic curvature balances the vacuum energy and the CC energy density as a consequence of the embedding. Thus, since the CC problem takes into account the fact that the gauge fields contributing to the vacuum energy are confined to the embedded space-time, the gravitational field, including the cosmological term is not. Therefore, a four-dimensional observer in the embedded space-time is able to perceive this difference through a conserved quantity built with the extrinsic curvature whose effect induces a warp effect in the embedded geometry. Interestingly, the extrinsic quantity $Q$ is a geometrical entity resulting from the extrinsic curvature and no prior \emph{ansatzes} were necessary. As a consequence, implications for nucleosynthesis epoch will be a subject of future research.

\end{document}